\documentclass{emulateapj} 
\usepackage{apjfonts}

\usepackage{amsmath}
\usepackage{verbatim} 
\usepackage{enumerate}
\usepackage{appendix}

\newcommand{\jdutc}{JD$_{\textrm{UTC}}$}
\newcommand{\jdtt}{JD$_{\textrm{TT}}$}
\newcommand{\jdtdb}{JD$_{\textrm{TDB}}$}

\newcommand{\hjdutc}{HJD$_{\textrm{UTC}}$}
\newcommand{\hjdutcp}{HJD$'_{\textrm{UTC}}$}
\newcommand{\hjdtt}{HJD$_{\textrm{TT}}$}
\newcommand{\hjdtdb}{HJD$_{\textrm{TDB}}$}

\newcommand{\bjdutc}{BJD$_{\textrm{UTC}}$}
\newcommand{\bjdutcp}{BJD$'_{\textrm{UTC}}$}
\newcommand{\bjdtt}{BJD$_{\textrm{TT}}$}
\newcommand{\bjdttp}{BJD$'_{\textrm{TT}}$}
\newcommand{\bjdtdb}{BJD$_{\textrm{TDB}}$}

\newcommand{\bjdtcb}{BJD$_{\textrm{TCB}}$}

\newcommand{\teph}{T$_{\textrm{eph}}$}
\newcommand{\jdct}{JD$_{\textrm{CT}}$}
\newcommand{\jdut}{JD$_{\textrm{UT}}$}

\shortauthors{EASTMAN ET AL}

\shorttitle{Achieving accurate Barycentric Julian Dates}

\begin{document}

\title{Achieving Better Than 1 Minute Accuracy in the Heliocentric
and Barycentric Julian Dates}

\author{Jason Eastman, Robert Siverd, and B. Scott Gaudi}

\affiliation{The Ohio State University, Columbus, OH 43210; jdeast@astronomy.ohio-state.edu}

\begin{abstract}

As the quality and quantity of astrophysical data continue to improve,
the precision with which certain astrophysical events can be timed
becomes limited not by the data themselves, but by the manner,
standard, and uniformity with which time itself is referenced. While
some areas of astronomy (most notably pulsar studies) have required
absolute time stamps with precisions of considerably better than 1
minute for many decades, recently new areas have crossed into this
regime. In particular, in the exoplanet community, we have found that
the (typically unspecified) time standards adopted by various groups
can differ by as much as a minute. Left uncorrected, this ambiguity
may be mistaken for transit timing variations and bias eccentricity
measurements. We argue that, since the commonly-used Julian Date, as
well as its heliocentric and barycentric counterparts, can be
specified in several time standards, it is imperative that their time
standards always be reported when accuracies of 1 minute are
required. We summarize the rationale behind our recommendation to
quote the site arrival time, in addition to using \bjdtdb, the
Barycentric Julian Date in the Barycentric Dynamical Time standard for
any astrophysical event. The \bjdtdb is the most practical absolute
time stamp for extra-terrestrial phenomena, and is ultimately limited
by the properties of the target system. We compile a general summary
of factors that must be considered in order to achieve timing
precisions ranging from 15 minutes to 1 $\mu$s. Finally, we provide
software tools that, in principal, allow one to calculate \bjdtdb \ to
a precision of 1 $\mu$s for any target from anywhere on Earth or from
any spacecraft.

\end{abstract}

\keywords{Barycentric, Heliocentric, Julian Date, UTC, TDB, TT, TDT,
TAI, light time} 
\maketitle

\section{Introduction}
Precise timing of astrophysical events is one of the fundamental tools
of astronomy, and is an important component of essentially every area
of study. There are two basic sources of uncertainty in timing: the
astrophysical data characterizing the event, and the time stamp with
which the event is referenced. Unfortunately, since the accuracy of
the time stamp is something that is often taken for granted, the
improvements in data are sometimes not accompanied (or are not
uniformly accompanied) by the requisite improvements in accuracy of
the time stamp used. This situation can lead to confusion, or even
spurious inferences.

Timing plays a particularly important role in the study of exoplanets.
Indeed, many of the ways in which exoplanets are discovered involve
the detection of transient or time-variable phenomena, including the
radial velocity, transit, microlensing, and astrometry techniques.
Furthermore, in some cases much can be learned about planetary systems
from the precise timing of these phenomena. As examples, the
measurement of terrestrial parallax in microlensing allows one to
infer the mass of the primary lens and so the planetary companion
\citep[e.g.,][]{gould2009}, and one can constrain the eccentricity of
transiting planets by comparing the times of primary transits and
secondary eclipses
\citep[e.g.,][]{deming05,charbonneau05,knutson07}. Possibly the most
promising application of timing in exoplanets, however, comes from
transit timing variations (TTVs). With an exquisitely periodic
phenomena like transiting planets, we will be able to measure many
effects using the departures from strict periodicity, such as the
gravitational perturbations from additional planets
\citep{miralda02,holman05,agol05}, trojans \citep{ford06b,ford07}, and
moons \citep{kipping09}, stellar quadrupoles \citep{miralda02}, tidal
deformations, general relativistic precession
\citep{jordan08,pal08a,heyl07}, orbital decay
\citep[e.g.,][]{sasselov03}, and proper motion \citep{rafikov09}.

Because of their great potential, TTVs have become the focus of many
groups. The typical data-limited transit timing precisions of most
observations are around 1 minute, with the best transit time precision
yet achieved of a few seconds \citep{pont07}. However, as discussed
above, accuracies of transit times are limited not only by the data
themselves, but also by the time stamp used. In order to make these
difficult measurements useful, it is critical that a time stamp be
used that is considerably more accurate than the uncertainty due to
the data themselves. Furthermore, since thorough characterization of
TTVs will require the use of all available data spanning many years
from several groups, this time stamp must be stable in the long term,
and all groups must clearly convey how it was calculated.

Unfortunately, we have discovered that, in the exoplanet community,
the Julian Date (JD) and its geocentric (GJD), heliocentric (HJD), and
barycentric (BJD) counterparts are currently being quoted in several
different, and often unspecified, time standards. In addition, the
site arrival time and its time standard is not quoted. This general
lack of homogeneity and specificity leaves quoted time stamps
ambiguous at the 1 minute level. More alarmingly, the most
commonly-used time standard, the Coordinated Universal Time (UTC), is
discontinuous and drifts with the addition of each leap second roughly
each year.

The pulsar community has solved the problem of precise timing well
beyond the level that is currently necessary for exoplanet studies,
and we can benefit from the techniques they have developed over the
past 40 years. In particular, their current state of the art program
({\tt TEMPO2}) models pulsar arrival times to 1 ns precision
\citep{hobbs06,edwards06}. This program is highly specialized and
generally cannot be applied outside of pulsar timing observations, but
many of the effects they consider are relevant to optical observers in
the exoplanet community.

In this article, we summarize the effects one must consider in order
to achieve timing accuracy of 1 $\mu$s -- well beyond the accuracy
that will likely be required by the exoplanet community for the
foreseeable future.

Section \ref{theory} provides the background required to understand
each of the effects that could change the arrival time of a
photon. They are listed in order of decreasing magnitude, so latter
subsections can be ignored for low-precision measurements.

Section \ref{practice} discusses the practical limitations to
achieving high-precision timing. We begin with the effects which may
cause errors that are comparable to or exceed the BJD
correction. These should be read and understood by everyone. We
continue with remaining effects, in order of decreasing magnitude,
which can be ignored for low-precision ($> 30$ ms) measurements. We
conclude \S \ 3 by listing additional effects, the errors due to which
are negligible ($< 1\mu$s).

We begin \S \ \ref{sec:calculating} by detailing the procedure one
must follow in order to calculate the \bjdtdb, which is designed to be
a useful reference for those already familiar with the concepts of
precision timing. In the latter part of this section, we describe our
particular IDL and web-based implementation of this procedure.

Lastly, in the Appendix, we discuss some of our specific findings
about the time stamps currently in use and how these are calculated
throughout the exoplanet community.

While we focus on the effects of timing on the optical/infrared
exoplanet community, timing precision of order 1 minute is necessary
for many other areas, such as the study of rapidly rotating white
dwarfs \citep{euchner06}. This article should be equally applicable in
such cases.

\section{Theoretical Timing Precision}
\label{theory}
The biggest source of confusion comes from the fact that time
standards and reference frames are independent from one another, even
though there are many overlapping concepts between the two. We will
use the following terminology: ``reference frame'' will refer to the
geometric location from which one could measure time -- different
reference frames differ by the light-travel time between them; ``time
standard'' will refer to the way a particular clock ticks and its
arbitrary zero point, as defined by international standards; and
``time stamp'' is the combination of the two, and determines the
timing accuracy of the event. 

The \bjdtdb, the time stamp we advocate, can be calculated using the
equation:

\begin{equation}
  \label{eq:bjd}
    BJD_{TDB} = JD_{UTC} + \Delta_{R\odot} + \Delta_{C} + \Delta_{S\odot} + \Delta_{E\odot},
\end{equation}

\noindent where \jdutc \ is the Julian Date in Coordinated Universal
Time; $\Delta_{R\odot}$ is the R{\o}mer Delay, discussed in \S \
\ref{sec:roemer}; $\Delta_{C}$ is the clock correction discussed in \S
\ \ref{sec:clock}; $\Delta_{S\odot}$ is the Shapiro delay discussed in
\S \ \ref{sec:shapiro}; and $\Delta_{E\odot}$ is the Einstein
delay, discussed in \S \ \ref{sec:einstein}.

The order of these terms is such that they are of decreasing
magnitude, so one need only keep the terms up to the precision
required. The timing precision required by current exoplanet studies
($\sim$ 1 s) requires only the terms up to and including $\Delta_{C}$.

Because future Solar System ephemerides may enable more precise
calculations of the arrival time at the Barycenter, or in order to
allow others to check that the original conversion was done accurately
enough for their purpose, the site arrival time (e.g., the \jdutc)
should always be quoted in addition to the \bjdtdb.

\subsection{Reference Frames: The R{\o}mer Delay}
\label{sec:roemer}
\begin{figure}
  \begin{center}
    \includegraphics[width=3.25in]{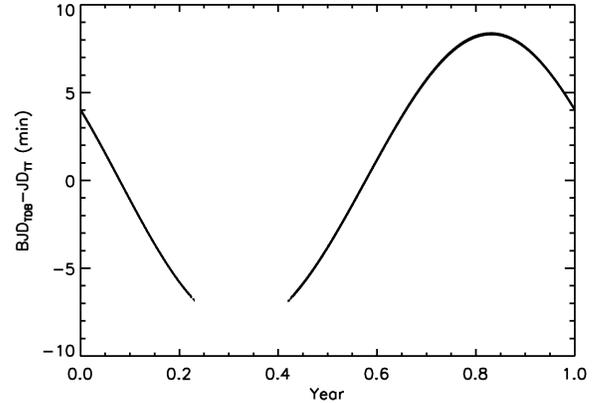} 
    \caption{Difference between \bjdtdb \ and the uncorrected \jdtt \
    (see text for definitions) over the course of a year. We plot the
    correction for a maximally-affected object on the ecliptic for an
    observer at latitude of zero degrees. We exclude all points where
    the object has an airmass greater than three and the Sun is higher
    than -12$^{\circ}$ in order to highlight observing biases.}
    \label{fig:bjdvjd}
  \end{center}
\end{figure}

Due to the finite speed of light, as the Earth travels in its orbit,
light from an astrophysical object may arrive early or be delayed by
as much as 8.3 minutes from the intrinsic time of the extraterrestrial
event. This is called the R{\o}mer delay, $\Delta_{R}$, in honor of
Ole R{\o}mer's demonstration that the speed of light is finite. Since
most observers cannot observe during daylight, a bias is introduced
and in practice the delay (as distinct from the early arrival time) is
only as much as 7 minutes, for a peak-to-peak variation of 15
minutes. Figure \ref{fig:bjdvjd} shows an example of this effect for a
maximally affected object on the ecliptic. In order to show the
observational bias, our example assumes the object is at 0$^\textrm{h}$ right
ascension and 0$^{\circ}$ declination. This curve shifts in phase with
ecliptic longitude and in amplitude with ecliptic latitude. We also
place our observer at the Earth's equator, but note that the asymmetry
will be larger at different latitudes.

The solution to this problem is to calculate the time when a photon
would have arrived at an inertial reference frame. This time delay is
the dot product of the unit vector from the observer to the object,
$\hat{n}$, and the vector from the origin of the new reference frame
to the observer, $\vec{r}$

\begin{equation}
  \label{eq:roemer}
  \Delta_{R\odot} = \frac{\vec{r} \cdot \hat{n}}{c},
\end{equation}

\noindent where $c$ is the speed of light and $\hat{n}$ can be written
in terms of its right ascension ($\alpha$) and declination ($\delta$),

\begin{equation}
  \label{eq:nhat}
  \hat{n} = 
  \begin{pmatrix}
    \cos(\delta)\cos(\alpha) \\
    \cos(\delta)\sin(\alpha) \\
    \sin(\delta)
  \end{pmatrix}.
\end{equation}
  
This equation is general as long as $\alpha$, $\delta$, and $\vec{r}$
are in the same coordinate system (e.g., Earth mean equator J2000) and
the object located at ($\alpha$, $\delta$) is infinitely far
away. Other forms of this equation in the literature assume that we
have the angular coordinates of the new origin or that the Earth and
the new origin are in the same plane \citep[e.g.,][]{binnendijk60,
henden82, hirshfeld97}, but we explain in \S \ \ref{sec:calculating}
why this form is most practical for calculating the delay.

\begin{figure}
  \begin{center}
    \includegraphics[width=3.25in]{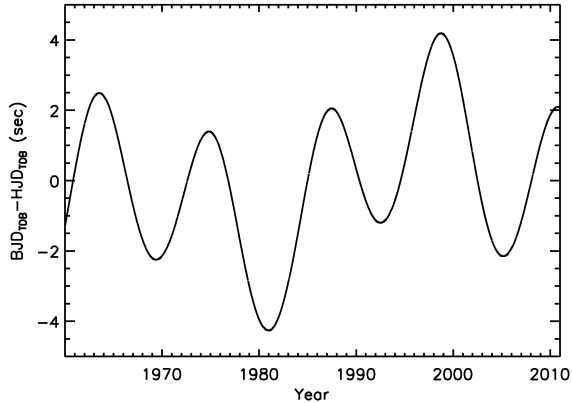} 
    \caption{Difference between the \bjdtdb \ and the \hjdtdb \ for a
      maximally-affected object on the ecliptic. The primary
      periodicity is due to Jupiter and the secondary periodicity is
      due to Saturn.}
    \label{fig:bjdvhjd}
  \end{center}
\end{figure}

The HJD, which uses the Sun as the origin of the new reference frame,
is only accurate to 8 s because of the acceleration of the Sun due
primarily to Jupiter and Saturn (Fig. \ref{fig:bjdvhjd}). It was
popular when people first began considering this effect because it is
relatively simple to calculate from tables without a computer
\citep{landolt72}, and remains popular because self-contained
algorithms exist to approximate it without any external tables
\citep[e.g.,][]{duffet89}. However, because the HJD is not useful when
accuracies of better than 8 s are needed, most of the algorithms in
use today use approximations that are only precise at the 1 s level,
and it becomes impossible to back out the original JD from the HJD
unless we know the exact algorithm used.

Because of these problems, the HJD was formally deprecated by
International Astronomical Union (IAU) Resolution A4 in 1991, in favor
of the BJD, a time referenced to the Solar System Barycenter (SSB).

The analogous correction to the R{\o}mer delay in our Solar System can
also be significant in the target system. We refer to this as
$\Delta_{R}$. For example, for transiting planets with $a \approx
0.06$ AU, $\Delta_{R}$ can be as large as 30 s. In general, the
position of the planet during primary transit has become the unspoken
standard reference frame for transiting planets, while the host star's
photosphere is the unspoken standard for Radial Velocity (RV) planets.

In theory, the timing would be much more stable in the target's
barycentric reference frame, but the accuracy with which we can
convert to this frame depends on the measurements of the system. Since
different observers may use different values as measurements improve,
quoting the JD in the frame of the target's barycenter may obfuscate
the long term reliability of timing. Therefore, we argue it is better
to quote Julian Date in the SSB reference frame, and correct for
$\Delta_{R}$ only when comparing observations at different phases in
the planet's orbit.

This $\Delta_{R}$ correction is not necessary for TTVs of the primary
transit, since the planet is always in the same phase. Nevertheless,
we should explicitly state the object's reference frame to avoid any
potential ambiguity, particularly when comparing any combination of
primary transits, secondary transits, RVs, and another primary transit
of a different planet in the same system, when it may not be obvious
which origin is being used.

For RV measurements, which are taken at many different phases, the
effect is much smaller and can generally be ignored because the star's
orbit around the barycenter is small. For a typical hot Jupiter,
(i.e., a Jupiter mass planet in a 3 day orbit around a solar mass
star), the maximum time difference in the RV signal (for an edge-on
orbit) is 20 ms, which would change the measured RV by $\sim 50 \
\mu$m s$^{-1}$. While planets farther out will cause a larger timing offset,
the difference in the measured RV is even smaller.

\subsection{Time Standards: The clock correction}
\label{sec:clock}
To be clear, the JD can be specified in many time standards
\citep{seidelmann92}, and while the IAU has made no explicit statement
regarding the allowed time standards of the GJD, HJD, or BJD, their
meaning in any given time standard is unambiguous. Unfortunately, they
have been specified in many standards, usually implicitly.

However, the particular time standard used affects how useful the time
stamp is as an absolute reference. We must be careful not to directly
compare BJDs or HJDs in different time standards, as each has
different offsets, periodic terms, and/or rates, which can introduce
systematic errors of over 1 minute. For this reason, it is critical
that any stated BJD or HJD also specify the time standard used when
one-minute accuracies are important, and the uncertainty of a time
that is quoted without a standard should be assumed to be at least 1
minute.

First, it may be useful to summarize the relevant standards of time:

\begin{description}

\item [Universal Time, UT1] -- Defined by the mean solar day, and so
drifts forward and backward with the speeding and slowing of the
Earth's rotation. Generally, it slows due to the tidal braking of the
Moon, though changes in the Earth's moment of inertia and complex
tidal interactions make its exact behavior unpredictable.  It is
rarely used directly in astronomy as a time reference, but we mention
it for context.

\item [International Atomic Time, TAI] -- Based on an average of
atomic clocks all corrected to run at the rate at the geoid at 0 K,
with 1 s equal to ``9,192,631,770 periods of the radiation
corresponding to the transition between the two hyperfine levels of
the ground state of the caesium 133 atom,'' as defined by Resolution 1
of the thirteenth meeting of the Conf\'{e}rence G\'{e}n\'{e}rale des
Poids et Mesures (CGPM) in 1967. This definition is based on the
duration of the Ephemeris Time second, which was previously defined as
1/31,556,925.9747 of the tropical year for 1900 January 0 at 12 hours
Ephemeris Time by Resolution 9 of the eleventh CGPM in 1960. TAI is
the fundamental basis for many other time standards, and is the
default time standard of the Sloan Digital Sky Survey.

\item [Coordinated Universal Time, UTC] -- Runs at the same rate as
TAI, except that it is not allowed to differ from UT1 by more than 0.9
s. Every 6 months, at the end of 31 December and 30 June, the
International Earth Rotation and Reference Systems Service
(IERS)\footnote{At http://www.iers.org.} may elect to add (or
subtract) a leap second to UTC in order to keep it within 0.9 s of
UT1. UTC is therefore discontinuous and drifts relative to TAI with
the addition of each leap second, which occur roughly once per
year. As of January 2009, the current number of leap seconds, $N$, is
34. The full table of leap seconds is available online and is
typically updated several months in advance of when an additional leap
second is to be added\footnote{At
ftp://maia.usno.navy.mil/ser7/tai-utc.dat.}. UTC is the current
international standard for broadcasting time. As a result, when a
modern, network-connected computer's clock is synchronized to a
Network Time Protocol (NTP) server, it will be in UTC. Thus, this is
the system of time most familiar to astronomers and non-astronomers
alike (modulo time zones and daylight savings time).

\item [Universal Time, UT] -- An imprecise term, and could mean UT1,
UTC, or any of several other variations. In general, such imprecise
language should be avoided, as the potential ambiguity is up to 1
s. In the context of a time stamp, it is likely UTC, but some people
may intentionally use UT to imply 1 s accuracy. While explicitness is
preferred (i.e., UTC $\pm$ 1 s), any time stamp quoted in UT should be
assumed to be uncertain at the 1 s level unless the time standard has
been independently verified.

\item [Terrestrial Time, TT(TAI)] -- A simple offset from TAI of
32.184 s released in real time from atomic clocks and never
altered. This offset is to maintain continuity between it and its
predecessor, the Ephemeris Time (ET).

\item [Terrestrial Time, TT(BIPM)] -- A more precise version of
TT(TAI). The International Bureau of Weights and Measures (BIPM)
reanalyzes TT(TAI) and computes a more precise scale to be used for
the most demanding timing applications. The current difference between
TT(TAI) and TT(BIPM) is $\sim30 \mu$s, and must be interpolated from a
table maintained by the BIPM and published online with a 1 month
delay\footnote{At ftp://tai.bipm.org/TFG/TT\%28BIPM\%29.}.

\item [Terrestrial Time, TT] -- Sometimes called Terrestrial Dynamical
Time (TDT), can refer to either TT(TAI) or TT(BIPM). From this point
on, we will not make the distinction, but when accuracies of better
than 30 $\mu$s are required, TT(BIPM) must be used.

\item [Barycentric Dynamical Time, TDB] -- Corrects TT for the
Einstein delay to the geocenter, $\Delta_{E\odot}$, which is the delay
due to time dilation and gravitational redshift from the motions of
the Sun and other bodies in the Solar System. The conversion from TT
to TDB cannot be written analytically, but is usually expressed as a
high-order series approximation \citep{irwin99}. The difference is a
predominantly a periodic correction with a peak-to-peak amplitude of
3.4 ms and a period of 1 yr. TDB was slightly modified by IAU
Resolution B3 in 2006, converging on the same definition as the JPL
Ephemeris Time, \teph, also called Coordinate Time (CT) in the JPL
ephemerides of Solar System objects.

\item [Barycentric Coordinate Time, TCB] -- Physically and
mathematically equivalent to the TDB as defined in 2006
\citep{standish98}, and differs only by an offset and rate of about
0.5 s yr$^{-1}$ due primarily to time dilation in the Sun's
gravitational potential. TDB and TCB were roughly equal at 1977 Jan
1.0 TAI, and now differ by about 16 s.

\end{description}

Caution must be always be exercised, however, as these definitions are
subject to change at any point, though usually with a few year's
notice.

Assuming the time is measured according to current definition of UTC,
then the clock correction, from UTC to TDB, can be written as the sum
of the corrections from UTC to TAI, TAI to TT\footnote{Ignoring the
TT(BIPM) correction.}, and TT to TDB:

\begin{equation}
  \label{eq:utc2tdb}
  \Delta_{C} = N + 32.184 s + (TDB - TT).
\end{equation}

Of course, if one wishes to express the BJD in another time standard
(or start with something other than UTC), the clock correction would
change accordingly. However, not every time standard is well-suited to
precise, astrophysical time stamps, and the use of any time standard
other than TDB should be viewed simply as an adequate approximation to
TDB.

Most readily available programs that calculate the time stamp assume
that the user has already applied the UTC-to-TT part of this clock
correction, which is often not true. We feel this assumption has
contributed the widespread confusion regarding time stamps.

As this last point is our primary motivation for writing this article,
we elaborate here on the effects of time standards on the reliability
of time stamps. For the sake of simplicity, we only discuss the
effects of time standards on the BJD. Each of these effects also
applies to the HJD, though the improvement in accuracy of the time
stamp is negligible compared to the accuracy of the HJD reference
frame for all but the UTC time standard.

\begin{figure}
  \begin{center}
    \includegraphics[width=3.25in]{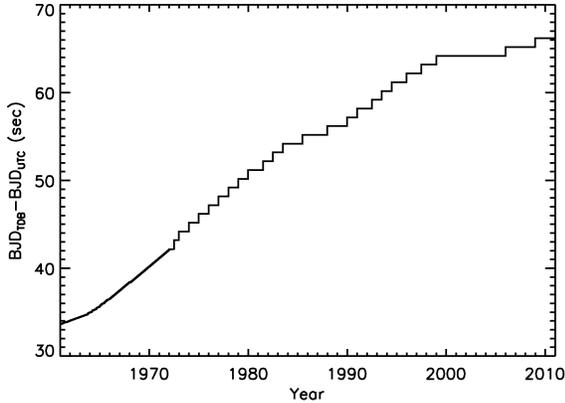} 
    \caption{Difference between the uniform \bjdtdb \ and the
    \bjdutc. It shows the discontinuities and slow drift in \bjdutc \
    due to the addition of leap seconds. Without correcting for these,
    relative timing between two reported values of \bjdutc \ can only
    be trusted over short time scales.}
    \label{fig:tdbvutc}
  \end{center}
\end{figure}

The least preferred, though most commonly used, time standard for the
BJD is UTC (\bjdutc), and is equivalent to ignoring $\Delta_{C}$
altogether. Because UTC is discontinuous and drifts with the addition
of each leap second, comparing two \bjdutc \ time stamps could result
in spurious differences if any leap seconds have been introduced
between observations. Therefore, 1 s timing accuracies cannot be
achieved using the \bjdutc \ over a span that straddles the addition
of one or more leap seconds (roughly 1 yr). Figure \ref{fig:tdbvutc}
shows the difference between the \bjdutc \ time stamp with the uniform
\bjdtdb \ time stamp (described below) from 1961 Jan 1, when UTC was
defined (though its definition has evolved over the years), to 2010
December 31, the furthest future date for which the value of UTC can
be accurately predicted at this writing.

BJD in TT (\bjdtt), which is equivalent to ignoring the (TDB-TT) term
in equation (\ref{eq:utc2tdb}), corrects for the discontinuity and
drift introduced by leap seconds and is appropriate for timing
accurate to 3.4 ms.

BJD in TDB (\bjdtdb) is usually the best time stamp to use in
practice, as it further corrects the \bjdtt \ for all known effects on
the motions, and therefore rates, of our atomic clocks. While \bjdtdb
\ is not perfect, any more accurate time stamp is unique for each
target.

BJD in TCB (\bjdtcb) corrects for the gravitational potential,
primarily from the Sun, which causes the clock to run slower than it
otherwise would. However, if one is concerned about effects of this
magnitude, the analogous correction of putting it in the gravitational
potential of each observed object is also required. Since these two
rates are small ($10^{-8}$), and opposite in sign, we believe it is
best to ignore \bjdtcb \ except perhaps as an intermediate step in
calculating the target-specific frame. Technically, the use of TCB was
recommended by IAU Resolution B1.3 in 2000. However, because of the
greater practicality of using TDB (see \S \ \ref{sec:calculating}),
and the drifting difference between TCB and TDB and TT, we believe its
use will only lead to confusion without any foreseeable benefit to the
exoplanet community.

\subsection{Shapiro Delay}
\label{sec:shapiro}
The Shapiro delay, $\Delta_{S}$ \citep{shapiro64}, is a general
relativistic effect in which light passing near a massive object is
delayed. For an object at an angle $\theta$ from the center of the
Sun, the Shapiro delay is

\begin{equation}
  \label{eq:shapiro}
  \Delta_{S\odot} = \frac{2GM_{\odot}}{c^3} \log(1-\cos{\theta}).
\end{equation}

\noindent This can be as large as 0.1 ms for observations at the limb
of the Sun, but for objects more than 30$^{\circ}$ from the Sun, the
correction is less than 20 $\mu$s.

There is also the analogous correction, $\Delta_{S}$, for the target
system. Similar to $\Delta_{R}$, $\Delta_{S}$ depends on the
measurements of the target, which may be refined over time. Therefore,
the time should generally be quoted without $\Delta_{S}$, but include
it when comparing times where this could be significant.

\subsection{Einstein Delay}
\label{sec:einstein}
As we discussed in \S \ \ref{sec:clock}, relativity dictates that the
motion of the observer influences the rate at which the observed clock
ticks. The use of TDB corrects for an observer moving with the
geocenter, but in reality we observe from the surface of the Earth or
from a satellite, for which there is an additional term to
$\Delta_{E\odot}$:

\begin{equation}
  \Delta_{E\odot} = \frac{\vec{r_{o}} \cdot \vec{v_{\Earth}}}{c^2}.
\end{equation}

\noindent Here, $\vec{r_{o}}$ is the location of the observer with
respect to the geocenter, and $\vec{v_{\Earth}}$ is the velocity of
the geocenter. Again, there is an analogous correction, $\Delta_{E}$,
for the target system, which should be ignored when quoting the time
but included when comparing times if necessary.

\section{Practical Considerations}
\label{practice}
Of course, the accuracy of the output time stamp is only as good as
our assumptions and the inputs needed to specify the time standard and
reference frame. Here we discuss their effects on the accuracy of the
time stamps. The first four subsections, through \S \
\ref{sec:planewave}, have no reasonable upper bound. In each case, the
accuracy of the inputs must be evaluated depending on the accuracy of
time stamp required. The later three subsections through \S \
\ref{sec:double} are organized in decreasing magnitude, all of which
can be ignored for accuracies no better than 21.3 ms. Finally, \S \
\ref{sec:negligible} discusses effects of less than 1 $\mu$s that we
have ignored.

\subsection{Coordinates}
\begin{figure}
  \begin{center}
    \includegraphics[width=3.25in]{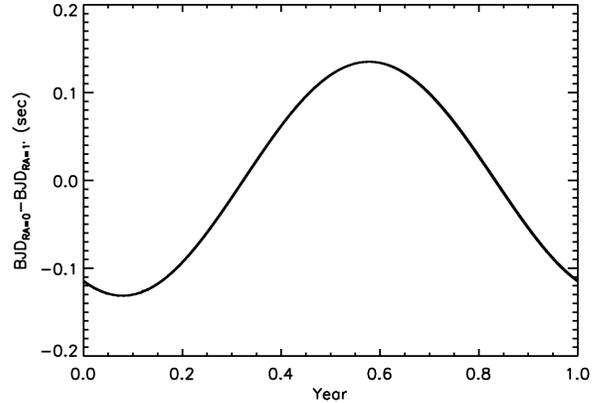} 
    \caption{Difference between the \bjdtdb \ calculated for an object
    at R.A. = 0$^{\rm{h}}$ and an object at R.A. = 0$^{\rm{h}}$
    0$^{\rm{m}}$ 4$^{\rm{s}}$ observed at the same time. This
    difference is as large as 200 s for a 10$^{\circ}$ offset.}
    \label{fig:bjdra}
  \end{center}
\end{figure}

An error of 1' in the position of the target amounts to a timing error
of as much as 0.28 s in the BJD (Fig. \ref{fig:bjdra}). Such error
would be common if the coordinates of the field center are used
instead of the specific object's coordinates. In particular, if doing
a survey, one may wish to assign the same BJD to all objects in a
given frame. However, with a 10$^{\circ}$ offset, which is possible
with some wide-field transit surveys, the error can be as large as 200
s for objects at the edge of the field. An error of 0.25'' will yield
1 ms timing offsets, and 0.25 mas accuracy is necessary for 1 $\mu$s
timing.

\subsection{Computer Clock}
\label{sec:clockacc}
The accuracy of a typical computer clock depends on its intrinsic
stability, the computer workload, its operating system, and the
reliability of the network connection. Older computers with a parallel
port CCD interface may produce unreliable timing because the clock may
slow or stop completely during CCD readout. Without any special
effort, a modern Windows machine with a network connection is accurate
to $\sim2$ seconds\footnote{At
http://support.microsoft.com/kb/939322.}, and with third-party
software like Dimension 4\footnote{At
http://www.thinkman.com/dimension4.}, we have found it to be stable to
0.1 s. An NTP-synchronized Linux machine is typically accurate to
$\sim50$ ms.

Of course, the stability of the clock only sets a lower limit on the
absolute accuracy of the time recorded in the FITS image header. NTP
synchronization attempts to measure and compensate for network
latency, but the accuracy of time stamps also depends on the
particular software package taking the image and the hardware it uses,
which is difficult to calibrate. Unless independently verified, the
time recorded in image headers should not be trusted to better than
0.25 s. However, various solutions exist for higher precision timing,
such as GPS-triggered shutters.

In particular, it is worth emphasizing the 1 s error in the {\it
Hubble Space Telescope} (HST) clock and potential 6.5 s error in the
Kepler clock, described in more detail in the Appendix, both of which
have already achieved data-limited transit timing precisions of that
order \citep[e.g.,][]{pont07,kipping10}.

\subsection{Flux-weighted mean time of exposure}
When calculating the time of exposure of an image, we typically use
the time at midexposure. However, the precise time of exposure that
should be used is the flux-weighted mean time of exposure. The
magnitude of this error depends on the intrinsic stability of source,
the stability of the atmosphere, and the exposure time. In the
diabolical case of a cloud completely covering the object during one
half of the exposure, the error could be as large as half the exposure
time. For a typical Hot Jupiter that dims by 1\% over the course of a
15 minute ingress, the error introduced into the time stamp during
ingress or egress by using the mid exposure time is 0.25\% of the
exposure time -- 150 ms for a 1 minute exposure. Near the peak of some
high-magnification microlensing events \citep[i.e.,][]{gould2009}, the
flux may double in as little as 6 minutes. During such instances,
using the mid exposure time will result in a time stamp error of as
much as 1/4 of the exposure time -- 15 s for a 1 minute exposure.

\subsection{Plane-Wave Approximation}
\label{sec:planewave}
Equation (\ref{eq:roemer}) assumes the object is infinitely far away,
and therefore the incoming wavefronts are plane waves. In reality, the
wavefronts are spherical, which introduces a distance-dependent,
systematic error. The maximum error introduced by the plane-wave
approximation is 1000 s for the Moon\footnote{The maximum error for
the Moon is not because of the departure from the plane wave, but
because the plane-wave formalism can place the Moon on the wrong side
of the SSB.}, 100 s for the Main Asteroid Belt, 5 s for the Kuiper
Belt, 1 ms at the distance of Proxima Centauri, and 150 ns at the
distance to the Galactic Center.

The fully precise equation, assuming spherical wavefronts, is

\begin{equation}
  \label{eq:exactroemer}
  \Delta_{R\odot} = \frac{\lvert \vec{r} + \hat{n}d \rvert - d}{c},
\end{equation}

\noindent where $d$ is the distance from the observer to the
target. In such instances, the distances must be derived from precise
ephemerides for both the target and observer.

Although this form is generally applicable, it is not generally
practical because at large distances, double precision floating point
arithmetic cannot reliably recover the small difference between
$\lvert \vec{r} + \hat{n}d \rvert$ (the distance from the barycenter
to the target) and $d$.

To solve this problem, the pulsar community
\citep[e.g.,][]{lorimer08}\footnote{Note: the sign of the second term
in Lorimer 2008, eq. 9, should be - rather than +.} uses the two-term
Taylor expansion of equation (\ref{eq:exactroemer}) about $1/d=0$:

\begin{equation}
  \label{eq:roemercorr}
  \Delta_{R\odot} \approx \frac{\vec{r} \cdot \hat{n}}{c} + \frac{\vec{r} \cdot \vec{r} - (\vec{r} \cdot \hat{n})^2}{2cd}.
\end{equation}

\noindent One may recognize the first term as plane wave approximation
(eq. [\ref{eq:roemer}]). In practice, the accuracy of this two-term
approximation exceeds the accuracy of the ``exact'' calculation using
double precision at a distance of 10,000 AU (0.05 pc). At the distance
of Proxima Centauri, the accuracy of this approximation is at worst 1
ns.

However, this Taylor expansion is divergent when $d < \lvert \vec{r}
\rvert$. It should never be used for objects inside 1 AU and may still
be inadequate for other objects inside the Solar System. It has a
maximum error of 1 day for the Moon, 20 s for the Main Asteroid Belt,
and 40 ms for the Kuiper Belt. In these cases, using the exact formula
(eq. [\ref{eq:exactroemer}]) may be easiest.

Therefore, we recommend that for precise calculations of any Solar
System body, equation (\ref{eq:exactroemer}) should be used. For
better than 1 ms timing of any object outside the Solar System,
equation (\ref{eq:roemercorr}) should be used.

\subsection{Geocenter}

\begin{figure}
  \begin{center}
    \includegraphics[width=3.25in]{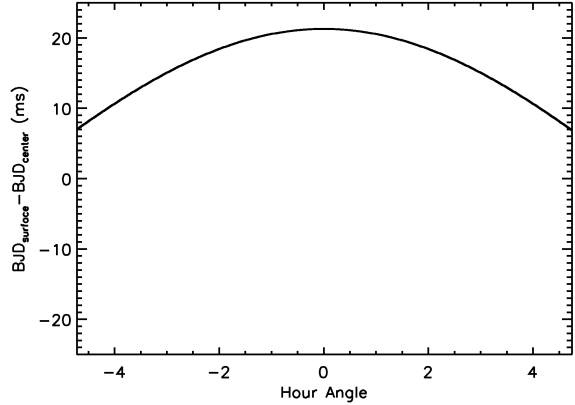} 
    \caption{Difference between the \bjdtdb \ calculated at the
    geocenter and at the precise location of the observer on the
    surface of the Earth. While geometrically, this effect will
    oscillate between $\pm 21.3$ ms with a period of 1 sidereal day,
    we exclude points when the Sun is above -12$^{\circ}$ and object
    is at $z > 3$, which introduces a large observational bias.}
    \label{fig:geocenter}
  \end{center}
\end{figure}

Most readily available time stamp calculators use the position of the
geocenter, rather than the location of the observer on the surface of
the Earth. Neglecting the light-travel time from the surface of the
Earth to the center introduces a 21.3 ms amplitude variation with a
period of 1 sidereal day. In practice, most observers can only observe
their targets at night, creating a systematic bias of between 8 ms and
21.3 ms (Fig. \ref{fig:geocenter}).

\subsection{\bjdutcp \ versus \bjdutc}

\begin{figure}
  \begin{center}
    \includegraphics[width=3.25in]{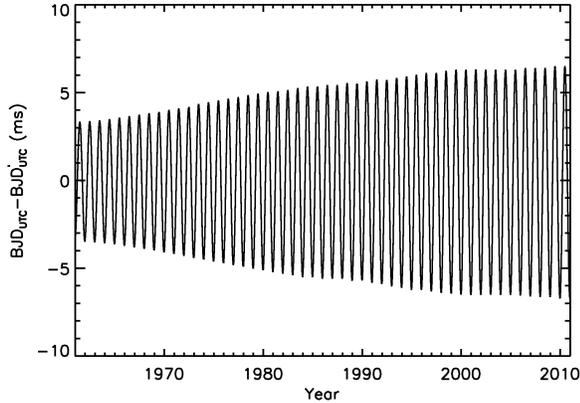} 
    \caption{Difference between the correctly calculated \bjdutc \ and
    the commonly calculated \bjdutcp \ using the positions of the
    Earth, planets, and Sun delayed by 32.184 + N s. It shows that the
    difference can safely be ignored for $\sim15$ ms precision, and
    therefore the approximate \bjdtt \ or \bjdtdb \ can be recovered
    from currently published \bjdutc \ simply by adding 32.184 + N s.}
    \label{fig:utcvutc}
  \end{center}
\end{figure}

Usually, when people calculate the \bjdutc, they neglect $\Delta_{C}$,
and input \jdutc \ for algorithms designed to take \jdtt. This
effectively uses the positions of the Earth, Sun, and planets offset
in time by 32.184 + N s to calculate the correction. When the \bjdutc
\ is calculated in this manner, we denote it as \bjdutcp. The correct
way to calculate the \bjdutc \ would be to first calculate the
\bjdtdb, then subtract the $\Delta_{C}$ correction\footnote{However,
the \bjdutc \ is a poor approximation to \bjdtdb \ and should not be
used.}. Figure \ref{fig:utcvutc} shows, for a maximally affected
object on the ecliptic, an example of the difference between the
\bjdutcp \ and the fully correct \bjdutc. Fortunately, this amounts to
at most a 13.4 ms difference (though growing with the UTC-TT
difference), which is below the precision of most clocks and the
geocentric correction that is usually ignored. Therefore, to an
accuracy of $\sim50$ ms, one can safely say \bjdtdb $\approx$ \bjdtt
$\approx$ \bjdutcp $+ \Delta_{C}$, making it easy to convert currently
published values of \bjdutcp \ to the superior \bjdtdb.

\subsection{Computer Precision}
\label{sec:double}
Representing JD as a double precision floating-point number limits the
accuracy to about 1 ms, and any operation done on the full JDs will be
even less accurate. Many programs require the use of a Reduced or
Modified Julian Date, and/or can return the JD to BJD offset in
seconds, but care must be taken at every step of the way never to
store the full JD as a double precision number if 1 ms precision is
required.

\subsection{Negligible ($< 1 \mu$s) effects}
\label{sec:negligible}
The Shapiro delay occurs for other bodies as well, but observations at
the limb of Jupiter only delay light by 200 ns.

Typical modern, commercial GPS units use the World Geodetic System
(WGS84), which is referenced to the International Terrestrial
Reference System (ITRS) with an error of about 15 m, which
amounts to a 50 ns error in the time stamp.

The index of refraction of the atmosphere is not exactly 1 and changes
with its composition, temperature, and pressure, which changes the
speed of light. However, the largest reasonable deviation due to this
effect is only tens of ns.

The pulsar community must specify a frequency-dependent dispersion
measure delay. At radio wavelengths (21 cm), the delay can be as much
as 1 s, but the dispersion delay contributes at less than 1 $\mu$s
shortward of $\sim300 \ \mu$m.

\section{Calculating the \bjdtdb}
\label{sec:calculating}
The most practical way to precisely calculate the \bjdtdb \ time stamp
is using JPL's DE405 ephemeris\footnote{At
ftp://ssd.jpl.nasa.gov/pub/eph/export/DE405/de405iom.ps.}. It contains
the position of thousands of bodies in the Solar System, including the
Sun, planets, spacecraft, moons, asteroids, and comets. It is oriented
to the International Celestial Reference Frame (ICRF), which is
consistent with the FK5 system at J2000.0 within the 50 mas error of
FK5, and has its origin at the SSB with its axes fixed with respect to
extragalactic objects \citep{arias95}. Therefore, it is recommended to
use the 3-space Cartesian coordinates retrieved from the JPL DE405
ephemeris directly with the J2000 object coordinates in equation
(\ref{eq:roemer}).

The following is an outline of the steps required to properly
calculate the \bjdtdb.

\begin{enumerate}

\item Calculate the midexposure time in \jdutc. Most FITS image
headers give DATE-OBS in UTC at the beginning of exposure. If high
precision is required ($<1$ s), read the caveats about clock precision
in \S \ \ref{sec:clockacc} carefully; depending on the sky conditions
or variability of the object, one may need to account for the
flux-weighted mean time of exposure.

\item Convert the midexposure time to \jdtdb \ by applying
$\Delta_{C}$ (eq. [\ref{eq:utc2tdb}]). For times accurate to 3.4 ms,
one can use the simpler \jdtt \ and calculate \bjdttp \ (using the
positions of planets delayed by the TT-TDB offset). The difference
between \bjdttp \ and \bjdtt \ is no more than $\sim$ 200 ns, which is
well below the precision of the \bjdtt. If better than 30 $\mu$s
precision is required, the TT(BIPM) - TT(TAI) offset must be applied.

\item Retrieve the JPL ephemeris of the observing station for the
times spanning the observing window. JPL's HORIZONS system\footnote{At
http://ssd.jpl.nasa.gov/?horizons.} is designed for this. To return
inputs for use with equation (\ref{eq:roemer}) and J2000 target
coordinates: select ``Vector Table'', the SSB as the coordinate
origin, and ``Earth mean equator and equinox of reference epoch'' for
the reference plane. This will return the Cartesian coordinates of the
observing station with respect to the SSB in the J2000 Earth mean
equator reference frame at steps as small as 1 minute in CT, which is
the same as TDB.

\item If the observing station is on Earth and better than 20 ms
timing is required, another ephemeris must be generated (from
HORIZONS) for the observer's position with respect to the non-rotating
geocenter, and added to the geocentric positions. We note that the
precise conversion from latitude, longitude, and elevation to the
Cartesian coordinates with respect to the non-rotating geocenter is
not trivial, and requires tables of measured precession and nutation
of the Earth.

\item Interpolate the positions of the observing station to each
midexposure time \jdtdb.

\item Input the interpolated X, Y, and Z positions of the observing
station, and the target's J2000 Earth mean equator coordinates into
equation (\ref{eq:roemer}). Depending on the distance to the target
and the precision required, equation (\ref{eq:exactroemer}) (and the
target's ephemeris) or equation (\ref{eq:roemercorr}) may be
required. One must be careful to use sufficiently accurate target
coordinates.

\item If greater than 0.1 ms precision is required, apply the Shapiro
correction (equation [\ref{eq:shapiro}]). An ephemeris of the Sun is
required, which can be generated from HORIZONS.

\item If greater than 1 $\mu$s precision is required, apply the
additional Einstein correction for the observing station's position
with respect to the geocenter. The geocentric velocity is required,
which can also be given by HORIZONS.

\end{enumerate}

\subsection{Our Code}
Our IDL code implementing this procedure is available
online\footnote{At
http://astroutils.astronomy.ohio-state.edu/time.}. It requires the
\jdutc \ at midexposure and target coordinates ($\alpha$, $\delta$) in
J2000 as inputs. We outline its procedure here. More explicit details,
as well as the calling procedure and dependencies, are commented
inside the code.

We compute $\Delta_{C}$ using Craig Markwardt's {\tt TAI\_UTC}
program\footnote{At
http://cow.physics.wisc.edu/\%7Ecraigm/idl/ephem.html.}  to read the
leap second table, and his {\tt TDB2TDT} program to compute the TT-TDB
correction, which uses a 791-term Fairhead and Bretagnon analytical
approximation to the full numerical integration, with an error of 23
ns \citep{fairhead90}.

Our code will automatically update its leap second table the first
time it runs after every January 1 or July 1, but this requires a
periodic Internet connection and the use of the {\tt wget} program. It
will terminate on failure to update, but this protection can be
bypassed for those that elect to (or have to) update their table by
hand.

By default, we ignore the $\sim30 \mu$s TT(BIPM) - TT(TAI) correction,
which would require a constant Internet connection, would not apply to
data acquired in the previous month, and is likely negligible for most
applications. However, our code can optionally correct for it if an
up-to-date file is supplied.

To read and interpolate the ephemeris from JPL, we use Craig
Markwardt's routines {\tt JPLEPHREAD} and {\tt JPLEPHINTERP} for the
Earth, Sun, and other planets. If the observing station is
space-borne, the smaller ephemeris used with those programs does not
include satellites, so we use an Expect script to automate a telnet
session to the HORIZONS system and automatically retrieve the
ephemeris, which we quadratically interpolate to the desired times
using IDL's {\tt INTERPOL}. The accuracy of this interpolation depends
on how quickly the object's position is changing and the step size of
the ephemeris. HORIZONS can only return $\sim90,000$ data points per
query, so the smallest step size (1 minute) limits the calculation to
a range of 60 days. For the geocenter, a 100 minute step size is
sufficient for 60 ns accuracy, but, for example, a 2 minute step size
is required for 1 $\mu$s accuracy for the {\it HST} (though it is
still limited by its clock accuracy). We have found that a 10 minute
step size is adequate for 1 ms timing for most objects and allows a
range of nearly 2 yr.

If the observer is on the Earth, and the coordinates (latitude,
longitude, and elevation) are given, we correct for the additional
delay. If no observer-specific information is given, we assume the
observer is at the geocenter, and the result will be biased by
$\sim10$ ms (Fig. \ref{fig:geocenter}).

If the target's ephemeris can be returned by HORIZONS and its unique
name is given, we use our Expect script to generate its ephemeris too,
and calculate the exact $\Delta_{R\odot}$
(eq. [\ref{eq:exactroemer}]). If not, and instead the distance is
given, we use the two-term approximation to the spherical wave
solution (eq. [\ref{eq:roemercorr}]). Otherwise, we use the plane wave
approximation (eq. [\ref{eq:roemer}]).

Lastly, we include the Shapiro correction and the additional Einstein
correction due to the position of the observer with respect to the
geocenter, either from the surface of the Earth (if given the
coordinates), or the spacecraft. 

In the geocentric case, our code agrees with {\tt BARYCEN}\footnote{At
http://astro.uni-tuebingen.de/software/idl/aitlib/astro/barycen.pro.}
to 200 ns (peak to peak) and the authors of BARYCEN report that their
code agrees with {\tt FXBARY}\footnote{At
http://heasarc.nasa.gov/lheasoft/ftools/fhelp/fxbary.txt.} to 1
$\mu$s. The ephemeris we generate for a location on the surface of the
Earth agrees with HORIZONS to 20 nano-lt-s, and the geocentric BJDs we
calculate from HORIZONS ephemeris agree with the BJDs we calculate
using Craig Markwardt's routines within 10 ns.

The near-exact agreements between these methods is not surprising, and
do not necessarily indicate that they are accurate to better than 1
$\mu$s. Our code was inspired by {\tt BARYCEN} and both rely on Craig
Markwardt's routines (the difference comes from the fact that we index
the JPL ephemeris with \jdtdb \ instead of \jdtt), and all methods use
JPL's DE405 ephemeris.

The primary advantage of our code is that it includes the \jdutc \ to
\jdtt \ correction (but can optionally ignore it). The choice of
starting with \jdutc \ is a departure from what is typically done with
such time stamp calculators, but we feel this is a far more robust
starting point. The current confusion has shown that many assume
\jdutc \ as the starting point, which is likely due to a lack of
explicitness in the programs and/or unfamiliarity with various time
standards. Our hope is that people are unlikely to make the opposite
mistake (assume the input should be \jdtt \ instead of \jdutc) since
our code is very explicit and calculating the \jdutc \ is almost
always a trivial calculation from the DATE-OBS FITS header keyword.

Additionally, our code can easily correct for the observer's position
on the Earth or from a spacecraft, and can include the spherical wave
correction.

In order to schedule observations, even $\pm$ 10 minute precision is
generally good enough, and one can approximate \bjdtdb \ $\approx$
\jdutc; for more demanding observing schedules, we provide software to
iteratively calculate the reverse correction.

Along with the IDL source code, we provide a web-based interface to
our codes\footnote{At
http://astroutils.astronomy.ohio-state.edu/time/utc2bjd.html.}, though
not every feature is enabled. Specifically, it is limited to 1 ms
precision, can only do one target at a time, only does the plane wave
approximation, and is limited to 10,000 JDs at a time. Those with
applications for which these features are too limited should download
our source code and run it locally.

\section{Conclusion}

Timing of transient events is a powerful tool for characterizing many
astronomical phenomena. In the field of exoplanets in particular, the
search for variations in the times of primary transits and secondary
eclipses, or transit timing variations, is one of the most promising
new techniques for studying planetary systems.

The accuracy with which transit times, and indeed any transient
phenomenon, can be measured is limited not only by the data
themselves, but by the time stamp to which the transit time is
referenced. As the quality of transit timing data crosses the
threshold of 1 minute precision, the precise time standard and
reference frame in which event times are quoted becomes important.
Achieving uniform and accurate time stamps with accuracies of better
than 1 minute that can reliably be compared to one another requires
extraordinary care in both our techniques and our terminology. We have
found that the time standards adopted by various groups that measure
transit times can differ by as much as a minute, and are typically
left unspecified. As these ambiguities can be significant compared to
the timing precisions that are quoted, they may therefore lead to
spurious detections of transit timing variations or biased
eccentricity measurements.

Here we have summarized the effects one must consider in order to
achieve timing precision of 1 $\mu$s. We argue that the \bjdtdb \ is
nearly the ideal time stamp, being as reliable as any time stamp can
be without being unique to each target system. On the other hand,
\bjdutc \ and the HJD in any form should be avoided whenever
possible. Most importantly, we emphasize that the time standard should
always be explicitly stated. Any time stamp that is quoted without a
time standard should be assumed to be uncertain to at least 1
minute. Unless the time standards used in programs or algorithms have
been independently confirmed, one should avoid using ones that do not
precisely specify the input and output time standard.

In addition, the arrival time at the observing site along with its
time standard (e.g., \jdutc) should also be specified. This will
remove any ambiguity in the time stamp, allow others to apply improved
corrections should more precise ephemerides become available in the
future, and allow others to check that original conversion was done
accurately enough for their purpose.

Finally, we have written an IDL program for general use that
facilitates the use of \bjdtdb \ to an accuracy of 1 $\mu$s, provided
that the inputs are sufficiently precise, and we provide a web-based
interface to its most useful features.

\begin{acknowledgments}
We would like to thank Craig Markwardt for his fundamental routines
that make ours possible, the help desks at the various space
telescopes and at IERS for answering our questions, the anonymous
referee, Steve Allen, Richard Pogge, Joseph Harrington, Roberto Assef,
Andrew Becker, Mercedes Lopez-Morales, Christopher Campo, Drake
Deming, Ryan Hardy, Heather Knutson, Eric Agol, and Joshua Winn for
useful discussions, and Wayne Landsman and J\"{o}rn Wilms for managing
the IDL astronomy libraries.
\end{acknowledgments}

\appendix
We looked in detail at several readily available tools for the BJD/HJD
calculation, and have been in contact with many people in the
exoplanet community and the help desks for several major space
telescopes. We summarize our findings here to demonstrate how easily
errors of up to 1 minute can be introduced and to stress the
importance of specifying the time precisely. We caution the reader not
to trust our general findings for specific cases, but always to
confirm what has been done in each case where 1 minute timing accuracy
is required but the time standard has not been specified explicitly.

\subsection{Software}

Ground-based observers have typically used one of the following
methods to calculate the BJD. However, most FITS image headers give
the DATE-OBS and TIME-OBS keywords in UTC. We have found that most
people, when starting with \jdutc \ end up quoting \hjdutcp \ or
\bjdutcp.

\subsubsection{HORIZONS}
JPL's HORIZONS ephemeris calculator, which is used by many to
calculate the BJDs from space telescopes, and can be used to calculate
ground-based BJDs, returns the time in \jdct = \teph = \jdtdb when the
ephemeris type is ``Vector Table''. Any conversion that uses a
HORIZONS ephemeris in \jdtdb \ but indexes it with \jdutc, as had been
done by several people we spoke with, will calculate \bjdutcp, which
can be offset from the true \bjdutc \ by up to 13.4 ms (as shown in
Fig. \ref{fig:utcvutc}), and offset from the uniform \bjdtdb \ by more
than 1 minute (as shown in Fig. \ref{fig:tdbvutc}).

\subsubsection{IRAF}
IRAF's {\tt setjd} calculates the HJD, but calls for UT, which is
likely to be interpreted as \jdutc. In this case, it would calculate
the drifting quantity \hjdutcp. If TT were used instead, it would
calculate the \hjdtt, accurate to $\sim$8 s.

\subsubsection{IDL}
The IDL routines {\tt HELIO\_JD} (for HJD)\footnote{At
http://idlastro.gsfc.nasa.gov/ftp/pro/astro/helio\_jd.pro.}, from the
IDL Astronomy Library curated by Wayne Landsman, and {\tt BARYCEN}
(for BJD), from the Institut f\"{u}r Astronomie und Astrophysik IDL
Library, maintained by J\"{o}rn Wilms, both call for the GJD, which,
we remind the reader, can be specified in any time standard. Often,
this is interpreted as \jdutc, in which case they would calculate
\hjdutcp \ or \bjdutcp, respectively. If TT were used, they would
calculate the \hjdtt, accurate to $\sim$8 s, or \bjdtdb, accurate to
the geocentric correction (21.3 ms).

\subsubsection{AXBARY}
NASA's High Energy Astrophysics Science Archive Research Center
(HEASARC) created the tools {\tt FXBARY} and later the improved
version {\tt FAXBARY}, both of which call {\tt AXBARY} to calculate
the BJD. Their documentation is precise and correct, but quite long
and may be difficult for the uninitiated to follow. Therefore, it
would not be surprising for users of {\tt AXBARY} to input either UTC
or TT, in which case they could generate either the \bjdutcp \ to the
accuracy of the leap seconds or the \bjdtdb \ to the accuracy of the
geocentric correction (21.3 ms).

\subsubsection{Online Tools}
Currently, common Google results turn up various applets,
spreadsheets, programs, or algorithms to calculate the HJD that
explicitly call for \jdutc \ or \jdut \ as an input. Unless explicitly
mentioned otherwise, it is usually safe to assume the time standard
used as input will be the time standard used throughout their
calculation. Thus, these algorithms and applets will very likely
calculate \hjdutcp. However, they are perfectly capable of calculating
\hjdtt \ if given \jdtt \ as an input.

\subsection{Space telescopes}

\subsubsection{EPOXI}
EPOXI has the mid-exposure time \bjdtdb \ in the header for the
intended pointing under the FITS header keyword KPKSSBJT. This can be
used directly, as long as one is careful about the intended target, so
it would be very surprising if a BJD from EPOXI was not \bjdtdb.

We recalculated the \bjdtdb \ using the HORIZONS ephemeris as
described in \S \ref{sec:calculating} and an example FITS header given
to us by the EPOXI help desk. In their example FITS header, they
pointed at the Moon, which is not infinitely far away, so we must use
equation \ref{eq:exactroemer} for $\Delta_{R\odot}$. With this method,
we agree with the KPKSSBJT header value to $\sim1$ ms, the limit of
the precision of the keyword.

We also redid the \bjdtdb \ calculation of HAT-P-4b as described in
the {\it Report on the Calibration of EPOXI spacecraft timing and
reduction to Barycentric Julian Date} of August 2009 by Hewagama et
al.. We find agreement with the quoted KPKSSBJT FITS header keyword to
47 ms. While this is much better than the 0.41 s difference calculated
by Hewagama et al. (a difference they attribute to ``cumulative
rounding limits'' in their method), we believe our method to be far
more precise. However, we could not obtain access to the original
headers and were unable to determine the source of the
discrepancy. Given the very good agreement with the calculation of the
Moon above, our best guess is that the target coordinates used by the
EPOXI pipeline differed from the published values for HAT-P-4b. The 47
ms difference could be explained by a $\sim27''$ discrepancy in R.A., a
$\sim45''$ discrepancy in declination, or some combination thereof.

\subsubsection{Chandra X-Ray Observatory}
Chandra stores their DATE-OBS keyword in TT. Their more precise TSTART
and TSTOP keywords are expressed in seconds after 1998 Jan 1, 00:00:00
TT. This departs from what is typically done, which may lead to
confusion, but it makes the conversion to a uniform time stamp much
more straightforward and less likely to drift by the leap
seconds. They provide extensive directions online to calculate the
\bjdtdb \ using {\tt AXBARY}, so it is likely that anyone using
Chandra who quotes a BJD is using \bjdtdb.

\subsubsection{Hubble Space Telescope}
The FITS headers of {\it HST} state that their DATE-OBS and TIME-OBS
keywords are UT. We contacted the {\it HST} help desk for
clarification, since UT is ambiguous. The {\it HST} help desk response
stated that their clock reports UTC accurate to $\sim$10 ms, but "due
to variabilities and quantization in the particular science
instruments' operations, the actual time light begins falling on the
detector is not known to better than about $\sim$1 second, $\pm$ 50\%
(rough estimate)." It is possible for the HST engineering team to
calibrate these variations, but they have limited resources and have
no plans to do so. It is thought that this error is some combination
of random and systematic errors, but the precise breakdown is unknown.

This potential $\sim$1 s systematic error may have important
implications for the reliability of the transit times quoted with HST
observations, most importantly, the 3 s error of the transit time of
HD189733b \citep{pont07}.

HST does not calculate the HJD or BJD at any point, leaving the
calculation up to each individual observer. Our experience with ground
based observers suggests that most people will end up quoting an
\hjdutcp \ or \bjdutcp.

\subsubsection{Kepler}
The Kepler Data Release Notes 2 describe how to calculate the BJD from
UTC, but do not include the correction to TDB. They mention the
HORIZONS ephemeris, but neglect to mention that its output time is in
CT, not UTC; thus it appears they calculate \bjdutcp, though we were
unable to confirm this. In addition, the time stamp uncertainty may be
much larger than typical, so it is worth quoting from the Kepler Data
Release Notes 5 (released 2010 June 4): The advice of the DAWG [Data
Analysis Working Group] is not to consider as scientifically
significant relative timing variations less than the read time (0.5 s)
or absolute timing accuracy better than one frame time (6.5 s) until
such time as the stability and accuracy of time stamps can be
documented to near the theoretical limit.

\subsubsection{Spitzer Space Telescope}
The {\it Spitzer} pipeline calculates the \hjdutc \ for the intended
pointing (presumably the target) at the end of the exposure, subtracts
the full exposure time, and records the result in the header as
HJD. Depending on the exposure time, this will produce roughly a 10 ms
effect similar to that shown in Figure \ref{fig:utcvutc}, and
depending on how close the intended pointing was to the object of
interest, may produce a $\sim 0.1$ s effect similar to Figure
\ref{fig:bjdra}. However, this effect is negligible compared to both
the $\sim$8 s accuracy of the HJD (Fig. \ref{fig:bjdvhjd}) and the
number of leap seconds that may have elapsed between observations
(Fig. \ref{fig:tdbvutc}).

One typically quotes the HJD at the midexposure time. Since {\it
Spitzer} quotes the \hjdutc \ at the beginning of the exposure, using
the unmodified {\it Spitzer} HJDs would produce a systematic offset of
half the exposure time, though experienced observers correct for this.

Also contributing to this confusion, the FITS header keyword UTCS\_OBS
is incorrectly documented. While the documentation states that it is
seconds after J2000 ET, it is actually seconds after January 1st, 2000
12:00 UTC + $N$ - 32. Therefore, trusting the documentation as is will
unwittingly lead to a difference of $N + 32.184$ s.

Most of the people we have asked opted to calculate their own BJD
using the HORIZONS ephemeris. However, they have typically quoted the
\bjdutcp.

\end{document}